\documentclass[prd,amsfonts,noshowpacs,nofootinbib]{revtex4}
\usepackage{amsmath,graphicx,hyperref,epsfig}

\begin{document}

\title{A note on the role of the boundary terms for the non-Gaussianity in general k-inflation}
\author{Frederico Arroja$^{1}$\footnote{arrojaf@ewha.ac.kr}and Takahiro
Tanaka$^{2}$\footnote{tanaka@yukawa.kyoto-u.ac.jp}}
\affiliation{{}$^{1}$Institute for the Early Universe, Ewha Womans
University, Seoul 120-750, South Korea\\ {}$^{2}$Yukawa Institute for
Theoretical Physics, Kyoto University, Kyoto 606-8502, Japan}

\begin{abstract}
In this short note we clarify the role of the boundary terms in the
 calculation of the leading order tree-level bispectrum in a fairly
 general minimally coupled single field inflationary model, where the
 inflaton's Lagrangian is a general function of the scalar field and its
 first derivatives. This includes k-inflation, DBI-inflation and
 standard kinetic term inflation as particular cases.
These boundary terms appear when simplifying the third order action by
 using integrations by parts. We perform the calculation in the comoving
 gauge obtaining explicitly all total time derivative interactions and
 show that a priori they cannot be neglected. The final result for the
 bispectrum is equal to the result present in the literature which was
 obtained using the field redefinition.
\end{abstract}


\date{\today}
\maketitle

\section{Introduction\label{sec:intro}}

The inflationary scenario has become the most compelling idea for what
really happened in the very early stages of our universe. The main
reason for this is that inflation is very successful at explaining the
problems of the old Big Bang model and it also provides us with a
mechanism to generate the primordial seed perturbations that later
evolved into the cosmic microwave background radiation anisotropies and
the large-scale structure of galaxies.

The most simple models of inflation predict a nearly Gaussian and nearly
scale invariant primordial perturbation in good agreement with recent
observations. However, a small amount of non-Gaussianity is still
allowed by the data. If such a small contribution were detected and
if it were of primordial origin, it would have profound implications for
inflationary models and our understanding of the very early
universe. Because of this reason, recently there has been a huge effort
to try to construct models that predict large (i.e. observable)
non-Gaussianity and calculate the higher-order correlation functions
like the bispectrum and the trispectrum. This search has been
productive. Many possibilities have been found, for example models with
non-canonical kinetic terms (like DBI-inflation, k-inflation,
ghost-inflation)~\cite{Creminelli:2003iq,Alishahiha:2004eh,Gruzinov:2004jx,Chen:2006nt,Huang:2006eha,Arroja:2008ga,Chen:2009bc,Arroja:2009pd,Huang:2010ab,Izumi:2010wm,Mizuno:2010ag,Burrage:2010cu},
multiple field models of inflation~\cite{Dvali:2003em,Enqvist:2004ey,Lyth:2005qk,Lyth:2005fi,Alabidi:2006hg,Sasaki:2006kq,Valiviita:2006mz,Sasaki:2008uc,Naruko:2008sq,Suyama:2008nt,Byrnes:2008wi,Byrnes:2008zz,Byrnes:2008zy,Cogollo:2008bi,Rodriguez:2008hy,Gao:2008dt,Langlois:2008vk,Langlois:2008wt,Langlois:2008qf,Arroja:2008yy,Chen:2009zp,Huang:2009xa,Huang:2009vk,Byrnes:2009qy,Langlois:2009ej,Mizuno:2009cv,Mizuno:2009mv,Gao:2009at,RenauxPetel:2009sj,Cai:2009hw,Kim:2010ud,Gao:2010xk},
temporary violations of the slow-roll conditions and small departures of
the initial vacuum state from the standard Bunch-Davies vacuum~\cite{Chen:2006xjb,Chen:2008wn,Hotchkiss:2009pj,Hannestad:2009yx,Flauger:2010ja,Chen:2010bk,Takamizu:2010xy}.
For recent reviews about these mechanisms to produce non-Gaussian
perturbations, see
Ref.~\cite{Koyama:2010xj,Chen:2010xk,Tanaka:2010km,Byrnes:2010em,Wands:2010af}.

In this note, we will be interested in the calculation of the bispectrum
for quite general models of k-inflation. Our conclusions will apply to
all k-inflation, DBI-inflation and standard kinetic term inflation
models. Our main goal is to clarify the role of the boundary terms in
the calculation of the bispectrum. These boundary interactions appear
when one does many integrations by parts to simplify the third order
action.

For the standard kinetic term inflation all these boundary terms have
been for the first time recently calculated in~\cite{Collins:2011mz}. However in all calculations of the bispectrum so
far these terms have been neglected and a field redefinition was used
instead. Maldacena~\cite{Maldacena:2002vr} states that these boundary
interactions are important and he takes them into account using field
redefinitions. We should point out one exception~\cite{Seery:2006tq},
where the authors working in the uniform curvature gauge and considering
standard kinetic term inflation show explicitly that the bispectrum can
be determined by using the third order action without the need to
redefine the field if the contribution from the boundary terms is
included.

In this work, we will perform an analogous calculation, but this time
for a general k-inflation model and working in the comoving gauge. In
this gauge, one works from the beginning with the variable of interest,
the comoving curvature perturbation. Naively, the third order action
does not seem to be suppressed by the slow-roll parameters contrarily to
the expected. However it has been shown~\cite{Maldacena:2002vr,Chen:2006nt} that after many integrations by
parts this suppression becomes evident. This procedure produces many
boundary terms which are the main focus of this work.

This paper is organized as follows. In section \ref{sec:model}, we will
introduce the model, the background spacetime and some useful
notation. In section \ref{sec:pert}, we will discuss linear
perturbations and present the solution for the mode functions. In
section \ref{sec:thirdorderaction}, we shall review the usual
calculation of the third order action and the bispectrum and obtain the
new boundary terms which we argue can be used to calculate the
bispectrum without the need to do a field redefinition. We also present
the calculation of the bispectrum produced by these time boundary
interactions. Section \ref{sec:conclusion} is devoted to the
conclusions.

\section{The model\label{sec:model}}

In this note, we will consider the class of models described by the
following Lagrangian
\begin{equation}
S=\frac{1}{2}\int d^4x\sqrt{-g}\left[M^2_{Pl}R+2P(X,\phi)\right],
\label{action}
\end{equation}
where $\phi$ is the inflaton field, $M_{Pl}$ is the reduced Planck mass,
$R$ is the Ricci scalar,
$X=-\frac{1}{2}g^{\mu\nu}\partial_{\mu}\phi\partial_{\nu}\phi$ is the
inflaton's kinetic energy and $g_{\mu\nu}$ is the metric
tensor. $P(X,\phi)$ denotes the inflaton's Lagrangian and we assume it
is a well-behaved function of its two variables. Throughout this work,
we use a system of units where the Planck constant $\hbar$, the speed of
light $c$ and the reduced Planck mass $M_{Pl}$ are set to unity. This
general Lagrangian includes as particular cases the DBI-inflation
model~\cite{Alishahiha:2004eh,Silverstein:2003hf} and the k-inflation
model~\cite{ArmendarizPicon:1999rj}.

We are interested in flat, homogeneous and isotropic
Friedmann-Lema\^{\i}tre-Robertson-Walker background universes
described by the line element
\begin{equation}
ds^2=-dt^2+a^2(t)\delta_{ij}dx^idx^j\,, \label{FLRW}
\end{equation}
where $a(t)$ is the scale factor.
The Friedmann equation and the
continuity equation read
\begin{equation}
3H^2=\rho, \quad
\dot{\rho}=-3H\left(\rho+P\right), \label{EinsteinEq}
\end{equation}
where dot denotes derivative with respect to cosmic time, the Hubble rate is $H=\dot{a}/a$, $\rho$ is the energy density of the
inflaton and it is given by
\begin{equation}
\rho=2XP_{,X}-P
,\label{energy}
\end{equation}
where $P_{,X}$ denotes the derivative of $P$ with respect to $X$.
It was shown in~\cite{Garriga:1999vw} that for this model the
speed of propagation of scalar perturbations (``speed of sound'')
is $c_s$ given by
\begin{equation}
c_s^2=\frac{P_{,X}}{\rho_{,X}}=\frac{P_{,X}}{P_{,X}+2XP_{,XX}}. \label{sound
speed}
\end{equation}
We define the slow-variation parameters, analogues of the
slow-roll parameters, as
\begin{equation}
\epsilon=-\frac{\dot{H}}{H^2}=\frac{XP_{,X}}{H^2}, \quad
\eta=\frac{\dot{\epsilon}}{\epsilon H}, \quad
s=\frac{\dot{c_s}}{c_sH}.
\end{equation}
We assume that the rate of change of the speed of sound is
small (as described by $s$) but $c_s$ is otherwise free to change
between zero and one.
It is convenient to introduce the following parameters that
describe the non-linear dependence of the Lagrangian on the
kinetic energy
\begin{equation}
\Sigma=XP_{,X}+2X^2P_{,XX}=\frac{H^2\epsilon}{c_s^2},\quad
\lambda=X^2P_{,XX}+\frac{2}{3}X^3P_{,XXX}. \label{sigmalambda}
\end{equation}
These parameters are related to the size of the bispectrum.

\section{Perturbations\label{sec:pert}}

In this section we will consider linear perturbations of the background
(\ref{FLRW}). There is a vast amount of works on linear perturbations,
see for example~\cite{Garriga:1999vw}.
It is convenient to use the Arnowitt, Deser and Misner (ADM) metric
formalism~\cite{Arnowitt:1960es}.
The ADM line element reads
\begin{equation}
ds^2=-N^2dt^2+h_{ij}\left(dx^i+N^idt\right)\left(dx^j+N^jdt\right),
\label{ADMmetricphi}
\end{equation}
where $N$ is the lapse function, $N^i$ is the shift vector and
$h_{ij}$ is the 3D metric.

The action (\ref{action}) becomes
\begin{equation}
S=\frac{1}{2}\int dt\,d^3x\sqrt{h}N\left({}^{(3)}\!R+2P\right)+
\frac{1}{2}\int dt\,d^3x\sqrt{h}N^{-1}\left(E_{ij}E^{ij}-E^2\right).\label{ADMaction}
\end{equation}
The tensor $E_{ij}$ is defined as
\begin{equation}
E_{ij}=\frac{1}{2}\left(\dot{h}_{ij}-\nabla_iN_j-\nabla_jN_i\right),
\end{equation}
and it is related to the extrinsic curvature of the spatial slices by
$K_{ij}=N^{-1}E_{ij}$. $\nabla_i$ is the covariant differentiation
with respect to $h_{ij}$ and all contra-variant indices in this section
are raised with $h_{ij}$ unless stated otherwise.

The Hamiltonian and momentum constraints are respectively
\begin{eqnarray}
{}^{(3)}\!R+2P-2\pi^2N^{-2}P_{,X}-N^{-2}\left(E_{ij}E^{ij}-E^2\right)&=&0,\nonumber\\
\nabla_j\left(N^{-1}E_i^j\right)-\nabla_i\left(N^{-1}E\right)&=&\pi
N^{-1}\nabla_i\phi P_{,X},\label{LMphi}
\end{eqnarray}
where $\pi$ is defined as $\pi\equiv \dot{\phi}-N^j\nabla_j\phi$.
We decompose the shift vector $N^i$ into scalar and intrinsic
vector parts as
$N_i=\tilde{N_i}+\partial_i\psi$,
where $\partial_i\tilde{N^i}=0$, here the index is raised with
$\delta_{ij}$.

On the comoving time-slices, the scalar field fluctuations vanish,
$\delta\phi=0$,
and the three-dimensional spatial metric $h_{ij}$ is perturbed as~\cite{Maldacena:2002vr}
\begin{eqnarray}
h_{ij}=a^2e^{2\zeta}\delta_{ij},
\end{eqnarray}
where $\zeta$ denotes the curvature perturbation on comoving slices and
tensor perturbations have been neglected because they do not contribute
to the tree-level scalar bispectrum.

We expand $N$ and $N^i$ in powers of the perturbation $\zeta$ as
\begin{eqnarray}
N=1+\alpha_1+\alpha_2+\cdots,\quad
\tilde{N_i}=\tilde{N_i}^{(1)}+\tilde{N_i}^{(2)}+\cdots,\quad
\psi=\psi_1+\psi_2+\cdots,
\end{eqnarray}
where $\alpha_n$, $\tilde{N_i}^{(n)}$ and $\psi_n$ are of order
$\zeta^n$.

Now, the strategy is to solve the constraint equations for the
lapse function and shift vector in terms of $\zeta$ and then plug
in the solutions in the expanded action up to the desired order. It
turns out that even for the third order action one does not need to use
the explicit solution for the constraints past first
order~\cite{Maldacena:2002vr,Chen:2006nt}.

At first order in $\zeta$, the solution for equations
(\ref{LMphi}) with a particular choice of boundary conditions at spatial infinity is~\cite{Maldacena:2002vr,Seery:2005wm,Chen:2006nt}
\begin{equation}
\alpha_1=\frac{\dot{\zeta}}{H}, \quad \tilde{N_i}^{(1)}=0, \quad
\psi_1=-\frac{\zeta}{H}+\chi, \quad
\partial^2\chi=a^2\frac{\epsilon}{c_s^2}\dot{\zeta}.\label{N1order}
\end{equation}

The second order action is then
\begin{equation}
S_2=\int dt\,d^3x\frac{a^3\epsilon}{c_s^2}\left(\dot \zeta^2-a^{-2}c_s^2(\partial\zeta)^2\right).
\label{2action}
\end{equation}

At leading order in slow-roll, $H$ is constant and the scale factor may
be approximated by that of a pure de Sitter universe $a=-1/(H\tau)$ and
$-Ht=\ln(-H\tau)$, where $\tau$ denotes conformal time. The solution of
the equation of motion derived from the previous action at leading order
in the slow-variation parameters and in Fourier space
is~\cite{Chen:2006nt}
\begin{equation}
\zeta(\tau,\mathbf{k})=\frac{iH}{\sqrt{4\epsilon c_sk^3}}(1+ikc_s\tau)e^{-ikc_s\tau}.\label{modefc}
\end{equation}

To quantize the curvature perturbation, we follow the standard procedure
in quantum field theory. We promote $\zeta$ to an operator that is
expanded in terms of creation and annihilation operators and mode
functions as
\begin{eqnarray}
\hat{\zeta}(\tau,\mathbf{k})=\zeta(\tau,\mathbf{k})a(\mathbf{k})
+\zeta^*(\tau,-\mathbf{k})a^{\dagger}(-\mathbf{k}),
\end{eqnarray}
where $a$ and $a^\dagger$ satisfy the usual commutation relation
$[a(\mathbf{k}),a^\dagger(\mathbf{k}')]=(2\pi)^3 \delta^{(3)}(\mathbf{k}-\mathbf{k}')$.

\section{The third order action and the bispectrum\label{sec:thirdorderaction}}

In this section, we will present the standard equations needed in the
calculation of the bispectrum of the primordial curvature perturbation
using the so-called in-in formalism~\cite{Schwinger:1960qe,Weinberg:2005vy}. In subsection
\ref{subsec:fieldredefinition} we shall consider the calculation using
the well-known field redefinition prescription and in subsection
\ref{subsec:boundaryterms} we will show that we can obtain the same
result if we include the contribution from the boundary terms that
appear when we simplify the action to the form (\ref{cubicaction}) by
using integrations by parts.

In order to use the machinery of the in-in formalism to compute the
tree-level three-point correlation function (or bispectrum) one needs to
calculate the cubic-order interaction Hamiltonian, see for example~\cite{Koyama:2010xj} for a review about this procedure.

\subsection{The field redefinition\label{subsec:fieldredefinition}}

The third order action, ignoring the many boundary terms, has been known
since the seminal work by Maldacena~\cite{Maldacena:2002vr} (for the
standard kinetic term case) and can be also found in~\cite{Seery:2005wm,Chen:2006nt} (for the model (\ref{action})), it reads

\begin{eqnarray}
S_3^{Reduced}&=&\int dt\,d^3x\bigg[
-a^3\left(\Sigma\left(1-\frac{1}{c_s^2}\right)+2\lambda\right)\frac{\dot\zeta^3}{H^3}
+\frac{a^3\epsilon}{c_s^4}\left(\epsilon-3+3c_s^2\right)\zeta\dot{\zeta}^2
+\frac{a\epsilon}{c_s^2}\left(\epsilon-2s+1-c_s^2\right)\zeta(\partial\zeta)^2
\nonumber \\ &&\qquad\qquad
-2a\frac{\epsilon}{c_s^2}\dot{\zeta}(\partial\zeta)(\partial \chi)
+\frac{a^3\epsilon}{2c_s^2}\frac{d}{dt}\left(\frac{\eta}{c_s^2}\right)\zeta^2\dot{\zeta}
+\frac{\epsilon}{2a}(\partial\zeta)(\partial
\chi) \partial^2 \chi +\frac{\epsilon}{4a}(\partial^2\zeta)(\partial
\chi)^2+ 2 f(\zeta)\left.\frac{\delta L}{\delta \zeta}\right|_1 \bigg],
\label{cubicaction}
\end{eqnarray}
where we should note that no slow-roll approximation has been made and we define
\begin{eqnarray}
\chi &=& \frac{a^2 \epsilon}{c_s^2} \partial^{-2} \dot \zeta, \label{shiftrel}\qquad
\frac{\delta
L}{\delta\zeta}\bigg |_1 = a
\left( \frac{d\partial^2\chi}{dt}+H\partial^2\chi
-\epsilon\partial^2\zeta \right), \\
f(\zeta)&=&\frac{\eta}{4c_s^2}\zeta^2+
 \frac{1}{c_s^2H}\zeta\dot\zeta+\frac{1}{4a^2H^2}\left[-(\partial\zeta)(\partial\zeta)+\partial^{-2}(\partial_i\partial_j(\partial_i\zeta\partial_j\zeta))\right]
+\frac{1}{2a^2H}\left[(\partial\zeta)(\partial\chi)-\partial^{-2}(\partial_i\partial_j(\partial_i\zeta\partial_j\chi))\right].
\nonumber\\
\label{redefinition}
\end{eqnarray}
In Eqs.~(\ref{shiftrel}) and (\ref{redefinition}), $\partial^{-2}$ denotes the inverse Laplacian and $\delta
L/\delta\zeta|_1$ denotes the variation of the quadratic action with
respect to the perturbation $\zeta$. The last term in
Eq.~(\ref{cubicaction}) can be absorbed by a field redefinition of
$\zeta$~\cite{Chen:2006nt},
\begin{eqnarray}
\zeta\rightarrow\zeta_n+f(\zeta_n). \label{fieldredef}
\end{eqnarray}
After this redefinition of variables, there appears a contribution
to the interaction Hamiltonian at cubic order coming from the
originally second-order action (See Eq.~(\ref{CV}) below).
This term is identical to the symmetric of the
last term in Eq.~(\ref{cubicaction}) up to a surface term arising
from the integration by parts. Neglecting such a surface term,
the interaction Hamiltonian in conformal time becomes
\begin{eqnarray}
H_{int}(\tau) &=&
-\int d^3x \Bigg[
-\left(\Sigma\left(1-\frac{1}{c_s^2}\right)+2\lambda\right)\frac{\zeta_n'^3}{H^3}
+\frac{a\epsilon}{c_s^4}\left(\epsilon-3+3c_s^2\right)\zeta_n\zeta_n'^2
+\frac{a\epsilon}{c_s^2}\left(\epsilon-2s+1-c_s^2\right)\zeta_n(\partial\zeta_n)^2
\nonumber \\ &&\qquad\qquad
-2\frac{\epsilon}{c_s^2}\zeta_n'(\partial\zeta_n)(\partial \chi_n)
+\frac{a\epsilon}{2c_s^2}\left(\frac{\eta}{c_s^2}\right)'\zeta_n^2\zeta_n'
+\frac{\epsilon}{2a}(\partial\zeta_n)(\partial
\chi_n) \partial^2 \chi_n +\frac{\epsilon}{4a}(\partial^2\zeta_n)(\partial
\chi_n)^2
\Bigg],
\label{Hint3}
\end{eqnarray}
where prime denotes derivative with respect to conformal time and $\chi_n=a\epsilon c_s^{-2}\partial^{-2}\zeta_n'$.

The field redefinition (\ref{fieldredef}) introduces extra terms in the three-point function as
\begin{eqnarray}
\langle\zeta(\mathbf{x}_1)\zeta(\mathbf{x}_2)\zeta(\mathbf{x}_3)\rangle
&=&\langle\zeta_n(\mathbf{x}_1)\zeta_n(\mathbf{x}_2)\zeta_n(\mathbf{x}_3)\rangle
\nonumber \\
&+&\frac{\eta}{2c_s^2}
\langle\zeta_n(\mathbf{x}_1) \zeta_n(\mathbf{x}_2)\rangle \langle\zeta_n(\mathbf{x}_1)\zeta_n(\mathbf{x}_3) \rangle
+ {\rm sym} + \mathcal{O}(\eta^2c_s^{-4} (P_\zeta)^3),
\label{3ptredef}
\end{eqnarray}
where ``sym'' denotes two terms that result from the preceding one by
symmetrizing with respect to $\mathbf{x}_1$, $\mathbf{x}_2$ and
$\mathbf{x}_3$.
The field redefinition (\ref{fieldredef}) includes several other terms,
however in the previous expression we only displayed the contribution of
the first term in (\ref{redefinition}). This is because the omitted
terms involve at least one derivative of $\zeta$ and they should vanish
when evaluated outside the horizon giving a negligible contribution to
(\ref{3ptredef}). For the reason mentioned above,
in this paper we approximate the function $f(\zeta)$ as
\begin{equation}
 f(\zeta)\approx \frac{\eta}{4c_s^2}\zeta^2.
\label{truncation}
\end{equation}
In the expression (\ref{3ptredef}),
the slow roll parameter $\eta$ has to be evaluated at the end of
inflation. This means that $\eta$ might become large depending on how
inflation ends which in turn would imply that the expansion in terms of
$\eta$ would cease to make sense. However in most cases one can safely
ignore the last term in Eq.~(\ref{3ptredef}), which is higher order in
$\eta$ because this is also greatly
suppressed by powers of the spectrum $P_\zeta$ which is measured to be
of order $10^{-10}$.

The tree-level three-point correlation function at the time $\tau_e$
after horizon exit is
\begin{equation}
\langle\Omega|\hat{\zeta}_n(\tau_e,\mathbf{k}_1)\hat{\zeta}_n(\tau_e,\mathbf{k}_2)\hat{\zeta}_n(\tau_e,\mathbf{k}_3)|\Omega\rangle=
-i\int_{-\infty}^{\tau_e} d\tilde\tau a\langle 0|
[
\hat{\zeta}_n(\tau_e,\mathbf{k}_1)\hat{\zeta}_n(\tau_e,\mathbf{k}_2)\hat{\zeta}_n(\tau_e,\mathbf{k}_3),{\hat{H}}_{int}(\tilde\tau)]
|0\rangle, \label{interaction}
\end{equation}
where $|\Omega\rangle$ and $|0\rangle$ denotes the interacting vacuum
and the free theory vacuum respectively. $[,]$ denotes the standard
commutator and the interaction Hamiltonian $\hat{H}_{int}$ is used to
evolve the free theory vacuum to the interaction vacuum at the time when
the three-point function is evaluated~\cite{Maldacena:2002vr}.

In Maldacena's calculation~\cite{Maldacena:2002vr} and in the following
ones~\cite{Seery:2005wm,Chen:2006nt} the last three terms in
Eq.~(\ref{Hint3})
which are higher-order in the slow-roll expansion were properly
neglected because these authors work at leading order in slow-roll. In
this work, we are only interested in the bispectrum produced by the
field redefinition, i.e. the second line of Eq.~(\ref{3ptredef}). In
Fourier space it reads~\cite{Chen:2006nt}
\begin{eqnarray}
\langle\Omega|\hat{\zeta}(0,\mathbf{k}_1)\hat{\zeta}(0,\mathbf{k}_2)\hat{\zeta}(0,\mathbf{k}_3)|\Omega\rangle
=(2\pi)^3\delta^{(3)}(\mathbf{K})\frac{H^4\eta}{32c_s^4\epsilon^2}\frac{k_1^3+k_2^3+k_3^3}{(k_1k_2k_3)^3},
\label{TPFfieldredefinition}
\end{eqnarray}
where $\mathbf{K}$ is defined as
$\mathbf{K}\equiv\mathbf{k}_1+\mathbf{k}_2+\mathbf{k}_3$.

\subsection{The boundary terms\label{subsec:boundaryterms}}

In the previous expression for the third order action
(\ref{cubicaction}) we omitted both time and spatial boundary terms as
in previous calculations. However the total action includes them and in
this subsection we will obtain these interactions explicitly. In the
following, total spatial derivative terms will be omitted because they do
not contribute to the three point function.
It is easy to see that the
interaction Hamiltonian of these terms is proportional to
$\mathbf{K}\equiv\mathbf{k}_1+\mathbf{k}_2+\mathbf{k}_3$ which has to
vanish because of momentum conservation imposed by the overall Dirac
delta function.

The third order action without neglecting the boundary terms is given by
\begin{eqnarray}
S_3^{Total}=S_3^{Reduced}+S_3^{Boundary},\label{cubicactiontotal}
\end{eqnarray}
and the explicit form of the boundary term is
\begin{eqnarray}
S_3^{Boundary}=
\int dt\,d^3x &\!\!\!\displaystyle\frac{d}{dt}\bigg[&\!\!\!\!
-9a^3H\zeta^3
+\frac{a}{H}\zeta(\partial\zeta)^2
-\frac{1}{4aH^3}(\partial\zeta)^2\partial^2\zeta
\nonumber\\
&&-\frac{a\epsilon}{c_s^2H}\zeta(\partial\zeta)^2
-\frac{\epsilon a^3}{c_s^4H}\zeta\dot\zeta^2
+\frac{1}{2aH^2}\zeta\left(\partial_i\partial_j\zeta\partial_i\partial_j\chi-\partial^2\zeta\partial^2\chi\right)
\nonumber\\
&&-\frac{\eta a}{2c_s^2}\zeta^2\partial^2\chi
-\frac{1}{2aH}\zeta\left(\partial_i\partial_j\chi\partial_i\partial_j\chi-\partial^2\chi\partial^2\chi\right)
\bigg]
,\label{Boundayaction}
\end{eqnarray}
where once again total spatial derivative terms were omitted.
One can calculate the three point functions coming from the individual
terms in this expression, directly using $\zeta$. The difference
in the third order action compared with the case of $\zeta_n$ is the
last term in (\ref{cubicaction}) and
$S_3^{Boundary}$.
The last term in (\ref{cubicaction}), $2 f(\zeta)(\delta L/\delta \zeta)|_1$
does not give any contribution because the factor
$(\delta L/\delta \zeta)|_1$ vanishes after substitution of the mode function.
The contributions from the different terms in $S_3^{Boundary}$ all vanish except for the
one coming from the first term in the third line (using the
leading order mode function and taking the limit $\tau_e\to 0$).

The contribution from the first term in the third line, i.e.
\begin{equation}
\int dt\,d^3x\frac{d}{dt}\left(-\frac{\eta\epsilon a^3}{2c_s^4}\zeta^2\dot\zeta\right),
\end{equation}
can be calculated using the in-in formalism to find
\begin{eqnarray}
\langle\Omega|&&\!\!\!\!\!\!\!\hat{\zeta}(\tau_e,\mathbf{k}_1)\hat{\zeta}(\tau_e,\mathbf{k}_2)\hat{\zeta}(\tau_e,\mathbf{k}_3)|\Omega\rangle\nonumber\\
&&=
(2\pi)^3\delta^{(3)}(\mathbf{K})
\Im\bigg[
\eta\epsilon c_s^{-4} a^2 \zeta(\tau_e,\mathbf{k}_1)\zeta(\tau_e,\mathbf{k}_2)\zeta(\tau_e,\mathbf{k}_3)
\zeta^*(\tau_e,\mathbf{k}_1)\zeta^*(\tau_e,\mathbf{k}_2)\zeta^{'*}(\tau_e,\mathbf{k}_3)
\bigg]\bigg |_{\tau_e\rightarrow0}+5\,\mathrm{perms.}
\nonumber\\
&&=(2\pi)^3\delta^{(3)}(\mathbf{K})\frac{H^4\eta}{32c_s^4\epsilon^2}\frac{k_1^3+k_2^3+k_3^3}{(k_1k_2k_3)^3},
\label{TPFboundary}
\end{eqnarray}
where we used the mode function solution (\ref{modefc}). This
bispectrum is of the so-called local type and exactly agrees with
Eq.~(\ref{TPFfieldredefinition}).
This result shows explicitly that the total time derivative terms cannot
a priori be neglected~\cite{Maldacena:2002vr,Seery:2006tq}. The total
bispectrum is the sum of the expression (\ref{TPFboundary}) (or
equivalently (\ref{TPFfieldredefinition}))
with the bispectrum
$\langle\zeta_n(0,\mathbf{k}_1)\zeta_n(0,\mathbf{k}_2)\zeta_n(0,\mathbf{k}_3)\rangle$,
which is calculated using Eq.~(\ref{interaction}).
This is the result that Chen \emph{et al.}~\cite{Chen:2006nt} got.
They performed the calculation by the field redefinition prescription
and used Wick's theorem as described in the preceding subsection.

In the approach of using the field redefinition explained in
the preceding section, we did not take into account the
boundary action $S_3^{Boundary}$.
If we perform the change of variables $\zeta=\zeta_n+f(\zeta_n)$
without neglecting the surface term,
it generates a third order action from the second order
 action (\ref{2action}) as
\begin{eqnarray}
S_3^{CV}&=&\int dt\,d^3x(-2)f(\zeta_n)\left.\frac{\delta L}{\delta\zeta_n}\right\vert_1
 +\int dt\,d^3x\frac{d}{dt}\left(2a^3\frac{\epsilon}{c_s^2}
  f(\zeta_n) \dot\zeta_n\right)
\nonumber\\
&\approx &\int dt\,d^3x(-2)f(\zeta_n)\left.\frac{\delta L}{\delta\zeta_n}\right\vert_1
+\int dt\,d^3x\frac{d}{dt}\left(\frac{a^3\epsilon\eta}{2c_s^4}\zeta_n^2 \dot\zeta_n\right),\label{CV}
\end{eqnarray}
where again total spatial derivative terms were omitted because they do
not contribute to the three point function,
and the approximate equality ``$\approx$'' means the truncation given in
Eq.~(\ref{truncation}), e.g. we neglect
total time derivatives terms that
do not contribute to the leading order three point function.
The first and second terms, respectively
eliminate the last terms in $S_3^{Reduced}$ proportional to the linear
equations of motion and the boundary term that
contributes to the bispectrum.

The boundary terms (\ref{Boundayaction}) are necessary to erase from the
action the terms with second order time derivative on $\zeta$ contained in the
last interaction in Eq.~(\ref{cubicaction}). Without these boundary
terms, the second order time derivative terms can remain in the action. These
problematic interactions from the point of view of the Hamiltonian
formalism are not present in the original action Eq.~(\ref{ADMaction}),
and they are not generated by the procedure of replacing the solution of
the constraints back into the action. In fact, these terms are generated
by the integrations by parts, and therefore the inclusion of the
boundary terms naturally takes care of them.

The calculations presented above clearly show that the boundary
terms can affect the result of the computation.
However, if the results of the computation depend on the choice of the total derivative terms in the action
$\int dt\,d^3x\, d{\cal B}(\zeta,\dot\zeta)/dt$,
one might be worried of how one can correctly choose ${\cal B}$.
The terms in ${\cal B}$ containing $\dot\zeta$
produce the second order time derivatives in the action.
As mentioned above, such problematic terms should vanish in total.
This condition completely determines the terms containing $\dot\zeta$ in
${\cal B}$.
In contrast, there is no criteria to choose the terms that
do not contain $\dot\zeta$. However, those terms do not contribute
to the equal-time expectation values of $\zeta$s.
In the path integral expression for the expectation value of an
operator ${\cal O}(t)$, adding such boundary terms given by ${\cal B}(\zeta)$
is equivalent to replacing the operator ${\cal O}$
with ${\cal O}'\equiv e^{-i\int d^3x {\cal B}(\zeta)}{\cal O}e^{i\int d^3x {\cal B}(\zeta)}$.
However, since ${\cal B}(\zeta)$ does not contain the conjugate
momentum of $\zeta$,
${\cal B}$ commutes with ${\cal O}$. Therefore ${\cal O}'$ reduces
to ${\cal O}$ and hence the results are not affected
by such boundary terms.

\section{Conclusion\label{sec:conclusion}}

We have computed the total time derivative interactions in the third
order action for the comoving curvature perturbation in general k-inflation. In
previous calculations of the bispectrum these boundary terms have been
ignored. In this note we have shown explicitly that a priori they are
important and should not be neglected freely. These boundary
interactions are
necessary to erase the terms with second-order time derivatives on
$\zeta$
in the action, generated by the integrations by parts. Total spatial
derivative terms can be safely ignored because their contribution for
the bispectrum is proportional to the sum of the three momentum vectors
which has to be zero due to momentum conservation.

From all the boundary terms that appear after many integrations by parts
in the action only one of them gives a non-zero bispectrum at leading
order in the slow-variation expansion. We have shown that the bispectrum
produced by this term is equal to the bispectrum produced in the usual
field redefinition prescription, thus our results agree with previous
results in the literature.

The main conclusion of this note is that in the calculation of the
bispectrum in general k-inflation one can ignore all the boundary terms that
appear when one simplifies the action but then one has to perform a
field redefinition to eliminate terms in the action that are
proportional to the first order equation of motion. On the other hand,
one might choose to keep all the boundary terms and calculate the
bispectrum using the usual method without the need to do the field
redefinition. We have shown that in the end the bispectrum of the
curvature perturbation is the same in both procedures.

An important lesson in computing the tree-level bispectrum
is that we should basically use the reduced action
written in terms of the physical variables that does not contain
second order time derivatives in total. If second order time derivatives are
contained, they must be eliminated by integration by parts.
Only when the action takes the canonical form without second order time
derivatives, one can use the expression for the third order action as
it is in the path integral expression.

\emph{Note added:} While we were writing up this work, the paper~\cite{Adshead:2011bw} appeared in the arXiv. They cite a paper in
preparation~\cite{Adshead:2011} where the authors also argue that the
inclusion of the boundary terms accounts for the terms introduced by the
field redefinition.

\emph{Note added in version 2:} Soon after this paper appeared on the arXiv, Ref. \cite{Adshead:2011} appeared as \cite{Burrage:2011hd}. Their section 3.1.2 contains similar, but independent, arguments on the role of the boundary terms. Some time after that, Ref. \cite{Rigopoulos:2011eq} appeared on-line and it also discusses the relation between field redefinitions, boundary terms and gauge transformations in the computation of the bispectrum.

\begin{acknowledgments}
FA would like to thank Kazuya Koyama, Kyung Kyu Kim and Misao Sasaki for
 interesting discussions. FA acknowledges the support by the World Class
 University grant no. R32-10130 through the National Research
 Foundation, Ministry of Education, Science and Technology
of Korea. TT is supported by JSPS Grant-in-Aid for Scientific Research
 (A) No. 21244033, the Global COE Program ``Next Generation of Physics,
 Spun from Universality and Emergence,'' and the Grant-in-Aid for
 Scientific Research on Innovative Areas Nos. 21111006
and 22111507 from the MEXT.
\end{acknowledgments}





\end{document}